\pdfoutput=1
\RequirePackage{ifpdf}
\ifpdf 
\documentclass[pdftex]{sigma}
\else
\documentclass{sigma}
\fi

\numberwithin{equation}{section}

\begin{document}
\allowdisplaybreaks

\newcommand{\arXivNumber}{1908.01530}

\renewcommand{\thefootnote}{}

\renewcommand{\PaperNumber}{003}

\FirstPageHeading

\ShortArticleName{On Complex Gamma-Function Integrals}

\ArticleName{On Complex Gamma-Function Integrals\footnote{This paper is a~contribution to the Special Issue on Elliptic Integrable Systems, Special Functions and Quantum Field Theory. The full collection is available at \href{https://www.emis.de/journals/SIGMA/elliptic-integrable-systems.html}{https://www.emis.de/journals/SIGMA/elliptic-integrable-systems.html}}}

\Author{Sergey \'E.~DERKACHOV~$^\dag$ and Alexander N.~MANASHOV~$^{\ddag\dag}$}

\AuthorNameForHeading{S.\'E.~Derkachov and A.N.~Manashov}

\Address{$^\dag$~St.~Petersburg Department of Steklov Mathematical Institute of Russian Academy of Sciences,\\
\hphantom{$^\dag$}~Fontanka 27, 191023 St.~Petersburg, Russia}
\EmailD{\href{mailto:derkach@pdmi.ras.ru}{derkach@pdmi.ras.ru}}

\Address{$^\ddag$~Institut f\"ur Theoretische Physik, Universit\"at Hamburg, D-22761 Hamburg, Germany}
\EmailD{\href{mailto:alexander.manashov@desy.de}{alexander.manashov@desy.de}}

\ArticleDates{Received October 15, 2019, in final form January 14, 2020; Published online January 18, 2020}

\Abstract{It was observed recently that relations between matrix elements of certain operators in the ${\rm SL}(2,\mathbb R)$ spin chain models take the form of multidimensional integrals derived by R.A.~Gustafson. The spin magnets with ${\rm SL}(2,\mathbb C)$ symmetry group and ${\rm L}_2(\mathbb C)$ as a local Hilbert space give rise to a new type of $\Gamma$-function integrals. In this work we present a direct calculation of two such integrals. We also analyse properties of these integrals and show that they comprise the star-triangle relations recently discussed in the literature. It is also shown that in the quasi-classical limit these integral identities are reduced to the duality relations for Dotsenko--Fateev integrals.}


\Classification{33C70; 81R12}

\renewcommand{\thefootnote}{\arabic{footnote}}
\setcounter{footnote}{0}

\section{Introduction}

The multidimensional integrals derived by R.A.~Gustafson~\cite{Gustafson92,Gustafson94} together with their $q$- and
elliptic analogues~\cite{Gustafson92,Gustafson94,Gustafson94-2,Spiridonov,MR2281166,Spiridonov06} play an important
role in different areas of physics and mathematics such as the theory of multi-variable orthogonal
polynomials~\cite{Stockman}, Selberg type integrals and constant term identities~\cite{AAR,Forrester-Warnaar}, and
supersymmetric dualities in quantum field theory~\cite{Spiridonov:2009za}. Recently, a new field~-- noncompact spin
magnets~-- was added to this list~\cite{Derkachov:2016dhc}.

Models of this type appear in gauge field theories and have been under intense investigations in the last two
decades, see references~\cite{Beisert:2010jr,Belitsky:2004cz}. The mathematical description of such systems is
well-developed and known as the quantum inverse scattering method(QISM)~\cite{MR1616371,Kulish:1981gi,Kulish:1981bi,Sklyanin:1988yz,Sklyanin:1991ss, STF}. Noncompact spin chains
have an infinite-dimensional local Hilbert space and most conveniently can be analysed within the separation of
variable (SoV) framework~\cite{Sklyanin:1991ss}. It was shown by Sklyanin that eigenfunctions of the entries of the
monodromy matrix provide a suitable basis for solving the spectral problem for the spin chain Hamiltonian. Building
such a~basis for a~generic spin chain is in itself a nontrivial problem and until recently the solution was available
only for the spin chains with a rank-$1$ symmetry group. For the noncompact magnets of interest the corresponding eigenfunctions are known explicitly~\cite{Belitsky:2014rba,MR1866770, Derkachov:2014gya}. It should be
mentioned here that recently there was a significant progress in constructing the SoV representation for compact
chains of higher ranks, see \cite{Cavaglia:2019pow, MR3868402,Gromov:2016itr,Kitanine:2016pvg,Kitanine:2018gki,Maillet:2018bim,Ryan:2018fyo}.

In the ${\rm SL}(2,\mathbb R)$ spin chin framework, Gustafson's integrals follow from identities for matrix elements
of the shift operator~\cite{Derkachov:2016dhc}.
 Extension of this analysis to the
$\mathrm{SL}(2,\mathbb{C})$ spin magnets leads to new integral identities~\cite{Derkachov:2016ucn,Derkachov:2017pvx}
which, up to the expected modification, are the exact replicas of the ${\rm SL}(2,\mathbb R)$ integrals. The analysis,
however, essentially depends on the completeness of the SoV representation. Proof of completeness is a rather
complicated problem, see, e.g., \cite{Derkachov:2018lyz,Kozlowski:2014jka,Schrader18}. Completeness for the closed
${\rm SL}(2,\mathbb R)$ spin chain follows from that of the Toda chain~\cite{Kozlowski:2014jka} while for the
${\rm SL}(2,\mathbb C)$ magnets of the length $N>2$ it is still an open question. A fruitful strategy seems to be to
make use of Gustafson's integrals to prove completeness. To realize this one needs an independent derivation of the
corresponding integrals. One of the purposes of this work is to provide such a derivation. We note also that complex
gamma integrals were studied recently by V.F.~Molchanov and Yu.A.~Neretin~\cite{Neretin} and
Yu.A.~Neretin~\cite{2018arXiv181207341N,Neretin:2019xze}.

The paper is organized as follows: in Section~\ref{sect:two}, after setting the notations, we prove two integral
identities which are direct analogs of Gustafson integrals associated with the classical $\textrm{su}(N)$ and
$\textrm{sp}(N)$ Lie algebras~\cite{Gustafson94}. We analyze the analytic properties of these integrals in
Section~\ref{sect:l-cases} and derive two new integrals which are the ${\rm SL}(2,\mathbb C)$ versions of Gustafson
integrals~\cite{Gustafson92} generalizing the second Barnes lemma. In Section~\ref{sect:star-triangle} it will be shown
that the $N=1,2$ integrals take, after some rewriting, the form of the star-triangle relations derived
in~\cite{Bazhanov:2007vg,Kels:2013ola,Kels:2015bda} as special limits of the elliptic star-triangle
identity~\cite{Bazhanov:2010kz,Derkachov:2012iv,Spiridonov18,Spiridonov:2010em}. Section~\ref{section:quasi} is devoted
to the study of the quasi-classical limit of the $\Gamma$-integrals. We show that in this limit the integrals are
equivalent to a special case of the duality relation~\cite{Baseilhac:1998eq} for Dotsenko--Fateev (DF)
integrals~\cite{Dotsenko:1984ad}. In Section~\ref{sect:FLI} we present an elementary proof of the above duality
relation and give some evidence which suggest that similar duality relations hold for the $\Gamma$-integrals.
Section~\ref{sect:summary} is reserved for a summary.

\section{ Gamma integrals}\label{sect:two}

\subsection{Definitions and basic properties}\label{subsec:def}

Let $u$, $\bar u$ be a pair of complex numbers of the form
\begin{gather}\label{def:z}
u=\frac n2+\nu, \qquad \bar u=-\frac n2+\nu,
\end{gather}
where $n$ is an integer and $\nu$ is complex number. We will use the notations $[u] = u-\bar u=n$ for the discrete
part and $\langle u \rangle = \nu$ for the continuous part and put $\|u\|^2=-u\bar u= -\nu^2+n^2/4$ so that for
imaginary $\nu$, $\|u\|^2\geq 0$. The $\boldsymbol \Gamma$~function of the complex field $\mathbb C$~\cite{GelfandGraevRetakh04} is defined as
\begin{gather*}
{\boldsymbol{\Gamma}}(u,\bar u)=\frac{\Gamma(u)}{\Gamma(1-\bar u)}
=\frac{\Gamma(n/2+\nu)}{\Gamma(1+n/2-\nu)}=(-1)^n\frac{\Gamma(-n/2+\nu)}{\Gamma(1-n/2-\nu)} =(-1)^{[u]}\Gamma(\bar u, u).
\end{gather*}
In what follows we will, for brevity, display only the first argument of the $\boldsymbol \Gamma$ function, i.e.,
$\boldsymbol \Gamma(u)\equiv \boldsymbol \Gamma(u,\bar u)$. Hereafter the following functional relations will be useful
\begin{gather}\label{func:gamma}
\boldsymbol \Gamma(u)\boldsymbol \Gamma(1-u)=(-1)^{[u]}, \qquad
\boldsymbol \Gamma(u+1)=-u\bar u \boldsymbol \Gamma(u).
\end{gather}
The $\boldsymbol \Gamma$ function appears in the generalization of Gustafson's integrals to the complex case.

The corresponding integrals take the following form~\cite{Derkachov:2016ucn,Derkachov:2017pvx}
\begin{subequations}\label{Gustafson12}
\begin{gather}
\frac1{N!} \sum_{n_1,\dots,n_N\in\mathbb{Z}+\frac{\sigma}{2}}
\int_{-{\rm i}\infty}^{{\rm i}\infty}
\frac{\prod\limits_{m=1}^{N+1}\prod\limits_{k=1}^{N}
\boldsymbol{\Gamma}(z_m-u_k)\boldsymbol{\Gamma}(u_k+w_m)}{ \prod\limits_{m<j}
\boldsymbol{\Gamma}(u_m- u_j)\boldsymbol{\Gamma}(u_j- u_m)}
\frac{{\rm d}\nu_1}{2\pi {\rm i}}\cdots\frac{{\rm d}\nu_N}{2\pi {\rm i}}\nonumber\\
\qquad{} =\frac{\prod\limits_{k,j=1}^{N+1}\boldsymbol{\Gamma}(z_k+ w_j)}{
\boldsymbol{\Gamma}\left(\sum\limits_{k=1}^{N+1} (z_k+w_k) \right)},\label{gustafson1}\\
\frac1{2^{N}N!}\sum_{n_1,\dots,n_N\in\mathbb{Z}+\frac{\sigma}{2}}
\int_{-{\rm i}\infty}^{{\rm i}\infty}
\frac{ \prod\limits_{k=1}^{N} \prod\limits_{m=1}^{2N+2}
 \boldsymbol{\Gamma}(z_m\pm u_k)}
{\prod\limits_{k=1}^{N}{\boldsymbol{\Gamma}}(\pm 2u_k) \prod\limits_{ k<j} {\boldsymbol{\Gamma}} (\pm u_k\pm u_j)
}\frac{{\rm d}\nu_1}{2\pi {\rm i}}\cdots\frac{{\rm d}\nu_N}{2\pi {\rm i}}\nonumber\\
\qquad{} = \frac{\prod\limits_{j<k}{\boldsymbol{\Gamma}}(z_j + z_k)}{{\boldsymbol{\Gamma}}\left(\sum\limits_{k=1}^{2N+2} z_k\right)}, \label{gustafson2}
\end{gather}
\end{subequations}
where we put $\boldsymbol \Gamma(a\pm b)\equiv \boldsymbol \Gamma(a+b) \boldsymbol \Gamma(a-b)$, $\boldsymbol
\Gamma(\pm a\pm b)\equiv \boldsymbol \Gamma(a+b) \boldsymbol \Gamma(a-b)\Gamma(-a+b) \boldsymbol \Gamma(-a-b)$. We will
refer to the integrals in the first and second lines as $I_N^{(1)}$ and $I_N^{(2)}$, respectively.

The variables $u_k$, $w_m$, $z_m$ have the form~\eqref{def:z}
\begin{gather*}
u_r=\frac{n_r}2+ \nu_r, \qquad z_j=\frac{m_j}2+ x_j, \qquad w_m=\frac{\ell_m}2+ y_m
\end{gather*}
and similarly for the barred variables. However, $n_r$, $m_i$, $\ell_m$ are allowed take integer or half-integer values,
simultaneously. Accordingly, the sums in~\eqref{Gustafson12} go over integers ($\sigma=0$) or half-integers
($\sigma=1/2)$. For the first integral~\eqref{gustafson1} there is no difference between the integer/half-integer cases
since they are related by the change of variables: $[u_r], [w_m], [z_j] \mapsto [u_r]+1/2, [w_m]-1/2,[z_j]+1/2$ so that we will assume that the variables $[u_k]$, $[w_m]$, $[z_m]$ in the integral~\eqref{gustafson1} are integers.

The integration contours in \eqref{Gustafson12} separate the series of $``\pm"$ poles due to the $\boldsymbol \Gamma$ functions in the numerators. The poles are located at
\begin{gather}\label{poles-I}
\nu^{\text{I},+}_{rj}(p)=\frac12{|n_r-m_j|} +x_j +p, \qquad \nu^{\text{I},-}_{rj}(p)=-\frac12{|n_r+\ell_j|} -y_j -p,
\qquad p\geq 0,
\end{gather}
where $r\in\{1, N\}$, $j\in\{1,N+1\}$ for the first integral and
\begin{gather*}
\nu^{\text{II},+}_{rj}(p)=\frac12{|n_r-m_j|} +x_j +p, \qquad \nu^{\text{II},-}_{rj}(p)=-\frac12{|n_r+m_j|} -x_j -p,\qquad p\geq 0,
\end{gather*}
where $r\in\{1, N\}$, $j\in\{1,2N+2\}$, for the second one.

Let us discuss now the convergence properties of the integrals~\eqref{Gustafson12}. Since the integrands are meromorphic functions and contours of integration avoid poles
it is sufficient to analyse the region of large $u_r=n_r/2+ {\rm i}\nu_r$, $ \|u_r\|^2=\nu_r^2+n_r^2/4 \to\infty$, only.
With the help of \eqref{func:gamma} we simplify the denominators in the integrals~\eqref{Gustafson12} as follows
\begin{gather}\label{den1}
\prod_{1\leq i<k\leq N} \frac1{\boldsymbol{\Gamma}(u_i- u_k)\boldsymbol{\Gamma}(u_k- u_i)} = (-1)^{(N+1)\sum_k [u_k] } \prod_{1\leq i<k\leq N}\|u_i-u_k\|^2
\end{gather}
and
\begin{gather}
\prod_{k=1}^{N}\frac1{\boldsymbol{\Gamma}(\pm 2u_k)} \prod_{1\leq i<m\leq N}\frac1{\boldsymbol \Gamma (\pm u_i\pm u_m)}\nonumber\\
\qquad{} =\varkappa_N 4^{N} \prod_{k=1}^N \|u_k\|^2
\prod_{1\leq i<m\leq N}\|u_i-u_m\|^2\|u_i+u_m\|^2,\label{den2}
\end{gather}
where $\varkappa_N=1$ for the integer case and $\varkappa_N=(-1)^{N(N+1)/2}$ for the half-integer case. Finally, taking
into account that for large $u$
\begin{gather}
\boldsymbol \Gamma(z -u) \boldsymbol \Gamma(u+w) =
(-1)^{[z-u]}\frac{\boldsymbol{\Gamma}(u+w)}{\boldsymbol{\Gamma}(u-z)} \nonumber\\
\hphantom{\boldsymbol \Gamma(z -u) \boldsymbol \Gamma(u+w)}{} = (-1)^{[z-u]}
 u^{z+w-1} (-\bar u)^{\bar z+\bar w -1}\bigl(1+O\left(1/\|u\|\right)\bigl) \label{large-u-as}
\end{gather}
we conclude that the integrals~\eqref{Gustafson12} converge absolutely provided
\begin{gather}\label{convergence}
 \sum_{j=1}^{N+1} \operatorname{Re}(x_j+y_j) <1 \qquad \text{and} \qquad \sum_{j=1}^{2N+2} \operatorname{Re} (x_j) <1,
\end{gather}
respectively. From now on we assume that these conditions are satisfied.

\subsection{Determinant representation}\label{subsec:det}

In this subsection we present the integrals~\eqref{Gustafson12} as determinants of $1$-dimensional integrals. Such a representation will be useful in what follows.\footnote{The determinant representation for elliptic hypergeometric integrals was constructed in~\cite{RainSpiridonov}.} For its derivation let us denote by $\mathcal Q (u|z,w)$ the function
\begin{gather*}
\mathcal Q (u|z,w)={\prod_{k=1}^{N+1}(-1)^{[u]}\boldsymbol\Gamma(z_k-u)\boldsymbol\Gamma(u+w_k)}
\end{gather*}
and by ${\mathcal{Q}}_{ik}(z,w)$ its Mellin moments
\begin{gather}\label{MellinQ}
{\mathcal{Q}}_{ik}(z,w)=\int \mathcal Du u^{i-1} (-\bar u)^{k-1} \mathcal Q (u|z,w), \qquad i,k=1,\ldots,N.
\end{gather}
Here we introduced a short-hand notation for the integration measure
\begin{gather*}
\int\mathcal Du\equiv
\sum_{n=-\infty}^{\infty}\int_{-{\rm i}\infty}^{{\rm i}\infty}\frac{{\rm d}\nu}{2\pi {\rm i}} .
\end{gather*}
Let $\boldsymbol{\mathcal Q}_N(z,w)$ be the following $N\times N$ matrix constructed from the Mellin moments
 \begin{gather}\label{Q-matrix}
\boldsymbol{\mathcal Q}_N(z,w)=\begin{pmatrix}
{\mathcal{Q}}_{11}(z,w)&\cdots& {\mathcal{Q}}_{1N}(z,w)
\\
\vdots & \ddots &\vdots
\\
{\mathcal{Q}}_{N1}(z,w)&\cdots& {\mathcal{Q}}_{NN}(z,w)
\end{pmatrix}.
\end{gather}
Rewriting the product on the r.h.s.\ of equation~\eqref{den1} as the product of two Vandermonde determinants
\begin{gather*}
\prod_{1\leq i<k\leq N}\|u_i-u_k\|^2=\Delta(u) \Delta (-\bar u), \qquad \Delta(u)=\det_{1\leq i,j\leq N} u_j^{i-1}
\end{gather*}
and taking into account the symmetry of the integrand in~\eqref{gustafson1} with respect to the permutations $u_i\leftrightarrow u_j$ we bring the first integral into the determinant form
\begin{gather}\label{det-rep-1}
I_N^{(1)} =\int\mathcal Du_1\cdots \mathcal Du_N \Delta(-\bar u) \prod_{k=1}^N \mathcal Q (u_k|z,w) u_k^{k-1}=\det \boldsymbol{\mathcal Q}_N(z,w).
\end{gather}
Proceeding in a similar way one gets the determinant representation for the second integral as follows
\begin{gather*}
I_N^{(2)} =\varkappa_N \det \boldsymbol{\widetilde {\mathcal Q}}_N(z)=\varkappa_N \det \begin{vmatrix}
\widetilde{\mathcal{Q}}_{11}(z)&\cdots& \widetilde{\mathcal{Q}}_{1N}(z)\\
\vdots & \ddots &\vdots\\
\widetilde{\mathcal{Q}}_{N1}(z)&\cdots& \widetilde{\mathcal{Q}}_{NN}(z)
\end{vmatrix},
\end{gather*}
where $\varkappa_N$ is a phase factor, see equation~\eqref{den2},
\begin{gather*}
\widetilde {\mathcal Q}_{ik}(z)=2\int \mathcal{D}u u^{2i-1}(-\bar u)^{2k-1} \widetilde {\mathcal Q}(u,z) \qquad\text{and}
\qquad
\widetilde{ \mathcal Q}(u,z)=\prod_{j=1}^{2N+2} \boldsymbol\Gamma(z_j\pm u).
\end{gather*}
Let us note here that the conditions~\eqref{convergence} are equivalent to the requirement of absolute convergence of the Mellin moments $\mathcal Q_{NN}(z,w)$ and $\widetilde{\mathcal Q}_{NN}(z)$, respectively.

\subsection{Proof of identities~(\ref{Gustafson12})}\label{subsec:proof}

Calculating the integral~\eqref{gustafson1} we will assume that the parameters $z_k$, $w_k$ satisfy the conditions
\begin{gather}\label{zw-cond}
\sum_{k=1}^{N+1} \operatorname{Re}(z_k+w_k)<1 \qquad \text{and} \qquad \sum_{k=1}^{N+1} \operatorname{Re} (\bar z_k+\bar w_k)<1.
\end{gather}
These conditions imply the condition~\eqref{convergence}, but do not follow from it and will be removed at the end of the calculation.

By virtue of~\eqref{convergence} the integrals~\eqref{MellinQ} are all absolutely convergent. We evaluate the integral
over $\nu$ by closing the contour in the left half-plane and then computing the sum over the residues. Recalling that
$u=n/2+\nu$ and $\bar u=-n/2+\nu$ one gets
\begin{gather*}
M_{ik}(n) =\int^{{\rm i}\infty}_{-{\rm i}\infty} u^{i-1}(-\bar u)^{k-1} \mathcal Q(u|z,w) \frac{{\rm d}\nu}{2\pi {\rm i}}\notag\\
\hphantom{M_{ik}(n)}{} =
(-1)^{(N+1)n}\sum_{j=1}^{N+1} \sum_{p=0}^{\infty}\frac{(-1)^p}{p! \bar p_j!} (-w_{j}-p)^{i-1} (\bar w_{j} + \bar p_j)^{k-1}\notag\\
\hphantom{M_{ik}(n)=}{} \times
\prod_{m=1}^{N+1} \frac{\Gamma(z_m+w_{j}+p)}{\Gamma(1-\bar z_m-\bar w_{j}-\bar p_j)}
\prod_{m \neq j}^{N+1} \frac{\Gamma(w_m-w_{j}-p)}{\Gamma(1-\bar w_m+ \bar w_{j}+ \bar p_j)},
\end{gather*}
where $\bar p_j = p+n+\ell_j$.

The first observation is that since the summand vanishes for $\bar p_j < 0$, only the poles at
$\nu^{\mathrm{I},-}_{rj}(p)$ given in~\eqref{poles-I}, contribute to the integral and, second, that under the
assumptions~\eqref{zw-cond} the sum over $p$ converges uniformly on $n$. Therefore, to evaluate $\sum_n M_{ik}(n)$ we
swap the summation over $n$ and $p$, change the summation variable from $n$ to $\bar p_j$ and finally suppress the
index $j$ in~$\bar p_j$: $\bar p_j\mapsto \bar p$
\begin{gather*}
\sum_n M_{ik}(n)=\sum_{j=1}^{N+1}\sum_{p= 0}^{\infty} \sum_{n=-\infty}^{\infty}( \ldots )
\mapsto \sum_{j=1}^{N+1}\sum_{p= 0}^{\infty}
\sum_{\bar p_j=-\infty}^{\infty}( \ldots ) =\sum_{j=1}^{N+1}\sum_{p= 0}^{\infty}
\sum_{\bar p =0}^{\infty}( \ldots ).
\end{gather*}
The final answer can be written in the form
\begin{gather}
\mathcal Q_{ik}(z,w) =\sum_{j=1}^{N+1} (-1)^{(N+1)[w_j]}
\left(\prod_{m=1}^{N+1}\frac{\Gamma(z_m+w_{j})}{\Gamma(1-\bar z_m-\bar w_{j})}\right)\left(
\prod_{\substack{m=1\\ m\neq j}}^{ N+1}\frac{ \Gamma(w_m-w_{j})}{\Gamma(1-\bar w_m+\bar w_{j})}\right)\notag\\
\hphantom{\mathcal Q_{ik}(z,w)=}{} \times
\left(\sum_{p=0}^{\infty} (-w_{j}-p)^{i-1}\prod_{m=1}^{N+1}
\frac{(z_m+w_{j})_{p}}{(1-w_m+w_{j})_{p}}\right)\nonumber\\
\hphantom{\mathcal Q_{ik}(z,w)=}{} \times
\left(\sum_{\bar p=0}^{\infty}(\bar w_{j} + \bar p)^{k-1}\prod_{m=1}^{N+1}
\frac{(\bar z_m+\bar w_{j})_{\bar p}}{(1-\bar w_m+\bar w_{j})_{\bar p}}\right) , \label{Q-factorized}
\end{gather}
where $(a)_p$ is the Pochhammer symbol. Thus $\mathcal Q_{ik}=\sum\limits_{j=1}^{N+1} \mathcal Q_{ik}(j)= \sum\limits_{j=1}^{N+1} \mathcal Q_{i}(j)\bar{\mathcal Q}_k(j)$. Hence the determinant can be represented as follows
\begin{gather*}
\det\mathcal Q =\sum_{j_1=1}^{N+1}\cdots \sum_{j_N=1}^{N+1}
\det\|\mathcal Q_{ik}(j_k)\| =\sum_{\sigma\in S_{N+1}}
\det \| \mathcal Q_{ik}(j_{\sigma(k)}\|
\notag\\
\hphantom{\det\mathcal Q}{} =\sum_{\sigma\in S_{N+1}} \left(\prod_{k=1}^N\bar{\mathcal Q}_k(j_{\sigma(k)})\right)
\det\| \mathcal Q_{i}(j_{\sigma(k)})\|,
\end{gather*}
where we take into account that $\det\|Q_{ik}(j_k)\|=0$ whenever $j_{m}=j_n$ for $n\neq m$. Then making use of~\eqref{Q-factorized} one can bring~\eqref{det-rep-1} into the following form
\begin{gather*}
I_N^{(1)}
=\frac1{N!}\!\sum_{\sigma\in S_{N+1}}\! (-1)^{(N+1)\sum\limits_{s=1}^N
[w_{\sigma(s)}]}\prod_{k=1}^{N}\left(
\prod_{j=1}^{N+1}\frac{\Gamma(z_j+w_{\sigma(k)})}{\Gamma(1-\bar z_j-\bar w_{\sigma(k)})}
\!\prod_{\substack{m=1\\ m\neq \sigma(k)}}^{N+1}\!
 \frac{ \Gamma(w_m-w_{\sigma(k)})}{\Gamma(1-\bar w_m+\bar w_{\sigma(k)})}\!\right)
\\
\hphantom{I_N^{(1)} =}{}
\times
\sum_{p_1,\ldots,p_N= 0}^\infty
 \left(\prod_{k<m} (w_{\sigma(k)}+p_k-w_{\sigma(m)}-p_m)\right)\prod_{m=1}^{N+1} \prod_{k=1}^{N}
\frac{(z_m+w_{\sigma(k)})_{p_k}}{(1-w_m+w_{\sigma(k)})_{p_k}}
\notag\\
\hphantom{I_N^{(1)} =}{}
\times
\sum_{\bar p_1,\ldots,\bar p_N=0}^\infty
 \left(\prod_{k<m} (\bar w_{\sigma(m)}+\bar p_m-\bar w_{\sigma(k)}-\bar p_k)\right)
\prod_{m=1}^{N+1} \prod_{k=1}^{N}
\frac{(\bar z_m+\bar w_{\sigma(k)})_{\bar p_k}}{(1-\bar w_m+\bar w_{\sigma(k)})_{\bar p_k}} .
\end{gather*}
The infinite sums over $\{ p\}$, $\{\bar p\}$ can be evaluated by Milne's ${\rm U}(n)$ Gauss
summation~\cite{MILNE198859}, see also~\cite[equation~(5.8)]{Gustafson94} and~\cite{Schlosser},
\begin{gather}
\sum_{p_1,\ldots,p_N = 0}^\infty \left(\prod_{1\leq k<m\leq N} (\alpha_{\sigma(k)}+p_k-\alpha_{\sigma(m)}-p_m)\right)
\prod_{i=1}^{N+1} \prod_{k=1}^{N}\frac{(\beta_i+\alpha_{\sigma(k)})_{p_k}}{(1-\alpha_i+\alpha_{\sigma(k)})_{p_k}}\nonumber\\
 \qquad{}=\frac{\Gamma\left(1-\sum\limits_{k=1}^{N+1}(\alpha_k+\beta_k)\right)
\prod\limits_{i=1}^{N}\Gamma(1+\alpha_{\sigma(i)}-\alpha_{\sigma(N+1)})}
{\prod\limits_{i=1}^{N+1}\Gamma(1-\beta_i-\alpha_{\sigma(N+1)})}.\label{Milne}
\end{gather}
Using~\eqref{Milne} we obtain the following representation for the integral $I_N^{(1)}$
\begin{gather*}
I_N^{(1)} =\frac1{N!}\!\sum_{\sigma\in S_{N+1}} \!(-1)^{(N+1)\sum\limits_{s=1}^N
[w_{\sigma(s)}] }\prod_{k=1}^{N}\left(
\prod_{i=1}^{N+1}\frac{\Gamma(z_i+w_{\sigma(k)})}{\Gamma(1-\bar z_i-\bar w_{\sigma(k)})}
\!\prod_{\substack{m=1\\ m\neq \sigma(k)}}^{N+1}\!
 \frac{ \Gamma(w_m-w_{\sigma(k)})}{\Gamma(1-\bar w_m+\bar w_{\sigma(k)})}\!\right)\\
\hphantom{I_N^{(1)} =}{} \times
\prod_{1\leq m<j\leq N} (w_{\sigma(m)}-w_{\sigma(j)})
\Gamma\left(1-\sum_{k=1}^{N+1}(z_k+w_k)\right)
\frac{\prod\limits_{k=1}^{N}\Gamma(1+w_{\sigma(k)}-w_{\sigma(N+1)})}
{\prod\limits_{k=1}^{N+1} \Gamma(1-z_k-w_{\sigma(N+1)}))}\\
\hphantom{I_N^{(1)} =}{} \times \prod_{1\leq m<j\leq N}
(\bar w_{\sigma(j)}-\bar w_{\sigma(m)})
\Gamma\left(1-\sum_{k=1}^{N+1}(\bar z_k+\bar w_k)\right)
\frac{\prod\limits_{k=1}^{N}
\Gamma(1+\bar w_{\sigma(k)}-\bar w_{\sigma(N+1)})}{
\prod\limits_{k=1}^{N+1}
\Gamma(1-\bar z_k-\bar w_{\sigma(N+1)})}.
\end{gather*}
After some simplifications this can be written as
\begin{gather*}
I_N^{(1)}=\frac{\Gamma\left(1-\sum\limits_{k=1}^{N+1}(\bar z_k+\bar w_k)\right)}{\Gamma\left(\sum\limits_{k=1}^{N+1}( z_k+ w_k)\right)}
\prod_{i,k=1}^{N+1}\frac{\Gamma(z_i+w_k)}{\Gamma(1-\bar z_i-\bar w_k)}
\frac{ R_N}{N!\sin\pi\left(\sum\limits_{k=1}^{N+1}( z_k+ w_k)\right)},
\end{gather*}
where
\begin{gather}
R_N = \sum_{\sigma\in S_{N+1}}
\frac{\prod\limits_{k=1}^{N+1}
\sin\pi(z_{\sigma(k)}+w_{\sigma(N+1)})}
{\prod\limits_{k=1}^N\sin\pi(w_{\sigma(N+1)}-w_{\sigma(k)})}
\nonumber\\
\hphantom{R_N =}{} \times(-1)^{(N+1)\sum\limits_{s=1}^N[w_{\sigma(s)}]}\prod_{1\leq k<j\leq N}
\frac{\sin\pi(\bar w_{\sigma(j)}-\bar w_{\sigma(k)})}{\sin\pi( w_{\sigma(j)}- w_{\sigma(k)})} .\label{R_N}
\end{gather}
Taking into account that $w_k-\bar w_k$ is an integer one finds that the last product in~\eqref{R_N} yields $(-1)^{(N-1)\sum\limits_{s=1}^N [w_{\sigma(s)}]}$ which cancels the second factor in~\eqref{R_N}. In the last step we use the \cite[Lemma~5.10]{Gustafson94} which states that
\begin{gather*}
\sum_{\sigma\in S_{N+1}}
\frac{\prod\limits_{k=1}^{N+1}
\sin\pi(\beta_{k}+\alpha_{\sigma(N+1)})}
{\prod\limits_{k=1}^N\sin\pi(\alpha_{\sigma(N+1)}-\alpha_{\sigma(k)})} =
N!\sin\pi \sum_{k=1}^{N+1}(\alpha_k+\beta_k).
\end{gather*}
It results in
\begin{gather*}
R_N=N!\sin\pi \sum_{k=1}^{N+1}( z_k+ w_k),
\end{gather*}
so that we get the required result for $I_N^{(1)}$
 \begin{gather*}
I_N^{(1)}=\frac{\prod\limits_{k,j=1}^{N+1}\boldsymbol{\Gamma}(z_k+ w_j)}{
\boldsymbol{\Gamma}\left(\sum\limits_{k=1}^{N+1} (z_k+w_k) \right)} .
\end{gather*}
Finally by analytic continuation in the $\nu_k$ the assumptions~\eqref{zw-cond} can be relaxed to the condition~\eqref{convergence}.

One sees that the crucial point in the proof of~\eqref{gustafson1} is the factorization of the double sum arising after evaluating the integral over $\nu$ by residue theorem into the product of two infinite sums, see equation~\eqref{Q-factorized}. We remark that this factorization was first noticed by Ismagi\-lov~\mbox{\cite{MR2265687,MR2354280}}, see also~\cite{Neretin:2019xze}. After this property is established the further analysis follows along the lines of~\cite{Gustafson94}.

The proof of the identity~\eqref{gustafson2} proceeds along the same lines so we will not go into details and only highlight the essential differences. First, we assume that the parameters $z_k$ satisfy the conditions
\begin{gather}\label{zz-cond}
\sum_{k=1}^{2N+2}\operatorname{Re}(z_k)<1 \qquad \text{and}\qquad \sum_{k=1}^{2N+2}\operatorname{Re}(\bar z_k)<1.
\end{gather}
After evaluating the integrals over $\nu$ using residue calculus, and carrying out some minor simplifications, we obtain
\begin{gather*}
I_N^{(2)} =\frac{2^N\varkappa_N}{N!}(-1)^N\sum_{\pi}
 \Biggl(\Biggl\{\sum_{y_{1},\ldots,y_{N}=0}^\infty
\left(\prod_{j=1}^N(z_{\pi(j)}+y_j)\prod_{1\leq i<j\leq N}\big(z_{\pi(i)}+y_i\pm(z_{\pi(j)}+y_j)\big)
\right)\\
\hphantom{I_N^{(2)} =}{} \times \prod_{k=1}^N \frac{(-1)^{y_k}}{y_k!} \Gamma(2 z_{\pi(k)}+y_k)\prod_{\substack{j=1\\j\neq \pi(k)}}^{2N+2}
{\Gamma(z_j\pm (z_{\pi(k)}+y_k))}\Biggr\}\notag\\
\hphantom{I_N^{(2)} =}{} \times
 \Biggl\{\sum_{\bar y_{1},\ldots,\bar y_{N}=0}^\infty
\left(\prod_{j=1}^N(\bar z_{\pi(j)}+\bar y_j)\prod_{1\leq i<j\leq N}
 \big(\bar z_{\pi(i)}+\bar y_i\pm(\bar z_{\pi(j)}+\bar y_j)\big)
\right) \notag\\
\hphantom{I_N^{(2)} =}{} \times \prod_{k=1}^N \frac{1}{\bar y_k!}
 \frac1{\Gamma(1-2\bar z_{\pi(k)}-\bar y_k)}\prod_{\substack{j=1\\j\neq \pi(k)}}^{2N+2}
\frac1{\Gamma(1-\bar z_j\pm (\bar z_{\pi(k)}+\bar y_k))}\Biggr\}\Biggr) ,
\end{gather*}
where $\pi$ is an injective map from $\{1,\ldots, N\}$ to $\{1,\ldots, 2N+2\}$. The sums over $y_j$, $\bar y_j$ can be
evaluated with the help of the following hypergeometric series summation formula, see \cite{MR1117471,Gustafson94}
\begin{gather*}
\sum_{y_1,\ldots,y_n=-\infty}^\infty \left\{\prod_{j=1}^n
 \frac{z_j+y_j}{z_j}\prod_{1\leq i<j\leq n}\frac{z_i+y_i\pm (z_j+y_j)}{z_i\pm z_j}
\prod_{i=1}^{2n+2}\prod_{j=1}^n\frac{\Gamma(w_i\pm(z_j+y_j))}{\Gamma(w_i\pm z_j)}
\right\}\\
\qquad{} = \frac{\Gamma\left(1-\sum\limits_{k=1}^{2n+2} w_i\right)\prod\limits_{i=1}^{2n+2}\prod\limits_{j=1}^n\Gamma(1-w_i\pm z_j)}{
\prod\limits_{j=1}^N\Gamma(1\pm 2 z_j){\prod\limits_{1\leq i<j\leq n} {\Gamma(1\pm z_i\pm z_j)}}
{\prod\limits_{1\leq i<j\leq 2n+2}{\Gamma(1-w_i-w_j) }}} .
\end{gather*}
Collecting all factors we obtain
\begin{gather*}
I_N^{(2)} = \frac{\prod\limits_{1\leq i<j\leq 2N+2}\boldsymbol\Gamma(z_i+z_j)}{\boldsymbol\Gamma\left( 1-\sum\limits_k z_k \right)} \times T_N ,
\end{gather*}
where
\begin{gather*}
T_N =\frac{(-1)^N}{2^{N} N!}\frac{\prod\limits_{1\leq i<j\leq 2N+2}\sin\pi(z_i+z_j)}{\sin\pi\left(\sum\limits_k z_k\right)}
 \sum_{\pi} \frac{\prod\limits_{1\leq i<j\leq N}
{\sin^2\pi(z_{\pi(i)}\pm z_{\pi(j)})} \prod\limits_{j=1}^N {\sin \pi (2 z_{\pi(j)})}}
{\prod\limits_{j=1}^N \prod\limits_{\substack{i=1\\ i\neq \pi(j)}}^{2N+2} {\sin\pi(z_i\pm z_{\pi(j)})}} .
\end{gather*}
One can show that $T_N=1$ by taking the limit $q\to 1$ of the identities in~\cite[equations~(7.11) and (7.12)]{Gustafson94}. Finally, the conditions~\eqref{zz-cond} can be relaxed to~\eqref{convergence} by analytic continuation in the $\nu_k$.

\section{Limiting cases}\label{sect:l-cases}

In this section we derive two more integrals which are replicas of the integrals~\cite[equations~(3.2) and (5.4)]{Gustafson92}. We show that in the complex case these integrals are intrinsically related to the integrals~\eqref{Gustafson12}. This property is not seen in the ${\rm SL}(2,\mathbb R)$ setup.

We start our analysis with the integral~\eqref{gustafson1} and introduce the variables, complex $\zeta$ and integer $\eta$, as follows
\begin{gather}\label{zeta-wz}
\zeta+\frac{\eta}2=\sum_{k=1}^{N+1} (z_k+w_k) , \qquad
\zeta-\frac{\eta}2=\sum_{k=1}^{N+1} (\bar z_k+\bar w_k).
\end{gather}
The r.h.s.\ of \eqref{gustafson1} is a meromorphic function of $\zeta$ with poles located at the points $\zeta_p=|1+\eta/2|+p$, $p\in \mathbb N_+$. For $\eta=0$ and $\zeta$ close to $1$ the r.h.s.\ of~\eqref{gustafson1} takes the form
\begin{gather}\label{Izetato1}
I_{N}^{(1)}=\frac1{1-\zeta}\prod_{k,j=1}^{N+1}\boldsymbol{\Gamma}(z_k+ w_j) +\cdots ,
\end{gather}
where $z_k$, $w_k$ obey the constraint $\sum\limits_{k=1}^{N+1} (z_k+w_k)=\sum\limits_{k=1}^{N+1} (\bar z_k+\bar w_k)=1$. At the same time, only the element $\mathcal Q_{NN}$ of the matrix~\eqref{Q-matrix} becomes singular at this point. The corresponding integral diverges at large $\nu$ and $n$ as $\zeta \to 1$. Indeed, taking into account
equation~\eqref{large-u-as} one finds
\begin{gather*}
{\mathcal{Q}}(u|z,w) = (-1)^{\sum_k [z_k]} u^{\zeta-N-1} (-\bar u)^{ \zeta-N-1}\big(1+O (1/{\|u\|} )\big).
\end{gather*}
Thus for $\zeta\to 1$
\begin{gather}\label{limitzetato1}
\mathcal Q_{N,N}(z,w) \simeq (-1)^{\sum_k [z_k]}\frac1\pi \int_{r>\Lambda} \frac{1}{ r^{4-2\zeta}} {\rm d}x{\rm d}y +\cdots
=\frac1{1-\zeta} (-1)^{\sum_k [z_k]}+\cdots
\end{gather}
and therefore
\begin{gather*}
I_{N}^{(1)}\underset{\zeta\to 1}{=} \mathcal Q_{NN}(z,w)\times \det\boldsymbol{\widehat{{\mathcal Q}}}_{N-1}(z,w) + \text{finite terms}.
\end{gather*}
Here $\boldsymbol{\widehat{{\mathcal Q}}}_{N-1} $ is the main $N-1$ minor of the matrix $\boldsymbol{\mathcal Q}_N$. Comparing the residues at $\zeta=1$ in equation~\eqref{limitzetato1} and \eqref{Izetato1} we obtain the following identity
\begin{gather}\label{newI}
\det\boldsymbol{\widehat{{\mathcal Q}}}_{N-1}(z,w)=
(-1)^{\sum_k [z_k]}\prod_{k,j=1}^{N+1}\boldsymbol{\Gamma}(z_k+ w_j) ,
\end{gather}
which holds provided $\sum\limits_{k=1}^{N+1} (z_k+w_k)=\sum\limits_{k=1}^{N+1} (\bar z_k+\bar w_k)=1$. Rewriting the l.h.s.\ of~\eqref{newI} in an integral form one obtains
\begin{gather}
\frac1{(N-1)!}\int \mathcal D u_1\cdots \mathcal Du_{N-1}
\prod_{p=1}^{N-1} (-1)^{n_p} \frac{\prod\limits_{j=1}^{N+1}\prod\limits_{k=1}^{N-1}\boldsymbol{\Gamma}(z_j-u_k)\boldsymbol{\Gamma}(u_k+w_j)}{
\prod\limits_{k<j} \boldsymbol{\Gamma}(u_k- u_j)\boldsymbol{\Gamma}(u_j- u_k)}\nonumber\\
\qquad{}=(-1)^{\sum_m [z_m]} \prod_{k,j=1}^{N+1}\boldsymbol{\Gamma}(z_k+ w_j).\label{gustafson1red}
\end{gather}
This integral is an analog of the integral~\cite[equation~(3.2)]{Gustafson92}. Indeed, replacing the variab\-le~$z_{N+1}$ by $z_{N+1}=1-\sum\limits_{k=1}^N z_k -\sum\limits_{m=1}^{N+1} w_m$ and using the relation~\eqref{func:gamma} one can bring equation~\eqref{gustafson1red} into the form which is a~replica of~\cite[equation~(3.2)]{Gustafson92}
\begin{gather}
\frac1{(N-1)!}
\int \mathcal D u_1\cdots \mathcal Du_{N-1}
\frac{\prod\limits_{k=1}^{N-1} \prod\limits_{m=1}^{N}\prod_{j=1}^{N+1}\boldsymbol{\Gamma}(z_m-u_k)\boldsymbol{\Gamma}(u_k+w_j)}{
\prod\limits_{m=1}^{N-1}\boldsymbol{\Gamma}(\gamma+u_m)
\prod\limits_{k<j} \boldsymbol{\Gamma}(u_k- u_j)\boldsymbol{\Gamma}(u_j- u_k)} \nonumber\\
\qquad{} =
\frac{\prod\limits_{k=1}^{N} \prod\limits_{j=1}^{N+1}\boldsymbol{\Gamma}(z_k+ w_j)}{\prod\limits_{j=1}^{N+1}\boldsymbol\Gamma(\gamma-w_j)},\label{gustafson-rform}
\end{gather}
where $\gamma=\sum\limits_{k=1}^N z_k + \sum\limits_{m=1}^{N+1} w_m$.

The analysis of the second integral, equation~\eqref{gustafson2}, proceeds along the same lines so we give only a brief account. Similar to \eqref{zeta-wz} we define variables, $\eta$ and~$\zeta$, by $\zeta+\eta/2=\sum\limits_{k=1}^{2N+2} z_k$. The l.h.s.\ and r.h.s.\ of equation~\eqref{gustafson2} have a pole at $\zeta=1$ (for $\eta=0$). Comparing the corresponding residues we get
\begin{gather}
\frac1{2^{N-1}(N-1)!} \int_\pm \mathcal D u_1\cdots \mathcal D u_{N-1}
\frac{ \prod\limits_{k=1}^{N-1} \prod\limits_{j=1}^{2N+2} \boldsymbol{\Gamma}(z_j\pm u_k)}
{\prod\limits_{k=1}^{N-1}{\boldsymbol{\Gamma}}(\pm 2u_k) \prod\limits_{j<k} {\boldsymbol{\Gamma}} (\pm u_k\pm u_j)}\nonumber\\
\qquad{} = \pm {\prod\limits_{1\leq j<k\leq 2N+2}{\boldsymbol{\Gamma}}(z_j + z_k)} ,\label{gustafson2red}
\end{gather}
where $\sum\limits_{k=1}^{2N+2} z_k=\sum\limits_{k=1}^{2N+2} \bar z_k=1$ and the subscript $\pm$ at the integral sign indicates that
the sum goes over either integer $n$ (plus) or half-integer~$n$ (minus). Introducing the variable
$\gamma=\sum\limits_{k=1}^{2N+1} z_k$, one can rewrite this integral in the form identical to the
integral~\cite[equation~(5.4)]{Gustafson92}
\begin{gather}
\frac{2^{1-N}}{(N-1)!}
\int_\pm \mathcal D u_1\cdots \mathcal Du_{N-1}
\frac{
 \prod\limits_{k=1}^{N-1} \prod\limits_{j=1}^{2N+1}
 \boldsymbol{\Gamma}(z_j\pm u_k)}
{\prod\limits_{k=1}^{N-1}\boldsymbol\Gamma(\gamma\pm u_k){\boldsymbol{\Gamma}}(\pm 2u_k)
\prod\limits_{j<k} {\boldsymbol{\Gamma}} (\pm u_k\pm u_j)}\nonumber\\
\qquad{} = \frac{\prod\limits_{1\leq j<k\leq 2N+1}{\boldsymbol{\Gamma}}(z_j + z_k)}{\prod\limits_{k=1}^{2N+3} \boldsymbol\Gamma(\gamma-z_k)}.\label{gustafson2-red2}
\end{gather}
Thus in the ${\rm SL}(2,\mathbb C)$ setup the integrals~\eqref{gustafson1red}, \eqref{gustafson-rform} and \eqref{gustafson2red}, \eqref{gustafson2-red2} are intrinsically related to the integrals~\eqref{Gustafson12}. For $N=2$ the relation (\ref{gustafson2-red2}) was derived by Sarkissian and Spiridonov~\cite{Spir}.

\section{Star-triangle relation}\label{sect:star-triangle}

In this section we show that the star-triangle relations with the Boltzmann weights given by a~product of $\Gamma$-functions~\cite{Bazhanov:2007vg,Kels:2013ola,Kels:2015bda} follow in a rather straightforward way from the integrals~\eqref{Gustafson12}. The star-triangle relation underlies exact solvability of various two-dimensional lattice models, see references~\cite{Baxter:1982zz,Bazhanov:2016ajm} for a review. Here we recall such a relation inherent to the noncompact~${\rm SL}(2,\mathbb C)$ spin chain magnets~\cite{MR1866770}.

Let $s_{\alpha}(z)\equiv s_{\alpha ,\bar\alpha}(z,\bar z)$ be a function of the complex variables~$z=x+{\rm i}y$, $\bar z\equiv z^*=x-{\rm i}y$,
\begin{gather*}
\boldsymbol s_\alpha(z) = [z]^{-\alpha}\equiv z^{-\alpha} \bar z^{-\bar\alpha}, \qquad [\alpha]=\alpha-\bar\alpha \in \mathbb Z.
\end{gather*}
The function $s_\alpha(z)$ is a single valued function on the complex plane and in the physics literature it is usually called a propagator.\footnote{Let us stress here that $[z]^\alpha$ denotes the power function while $[\alpha]$ without any superscript stands for the ``integer'' part of $\alpha$, $[\alpha]=\alpha-\bar\alpha$. We hope that this
slightly unfortunate notation does not lead to confusion.} It satisfies two identities:
\begin{itemize}\itemsep=0pt
\item the chain relation
\begin{gather}\label{chain-C}
\frac1\pi\int \boldsymbol
s_{\alpha_1} (z_1-z) \boldsymbol s_{\alpha_2}(z-z_2) {\rm d}^2 z =
\frac{\boldsymbol \Gamma(1-\alpha_1)\boldsymbol \Gamma(1-\alpha_2)}{\boldsymbol \Gamma(2-\alpha_1-\alpha_2)} \boldsymbol s_{\alpha_1+\alpha_2-1}(z_1-z_2),
\end{gather}
\item the star-triangle relation
\begin{gather}
\frac1\pi\int\prod_{k=1}^3 \boldsymbol s_{\alpha_k}(z_k-z) {\rm d}^2z \nonumber\\
\qquad{} =\left(\prod_{k=1}^3 \boldsymbol \Gamma(1-\alpha_k)\right)
\boldsymbol s_{1-\alpha_1}(z_2-z_3)\boldsymbol s_{1-\alpha_2}(z_3-z_1)\boldsymbol s_{1-\alpha_3}(z_1-z_2), \label{star-triangle-C}
\end{gather}
which holds provided $\alpha_1+\alpha_2+\alpha_3=\bar \alpha_1+\bar \alpha_2+\bar \alpha_3=2$.
\end{itemize}
In fact these two relations are equivalent: equation~\eqref{star-triangle-C} is reduced to equation~\eqref{chain-C} in the limit $z_3\to \infty$ and, vice versa, equation~\eqref{star-triangle-C} can be derived from equation~\eqref{chain-C} by a~${\rm SL}(2,\mathbb C)$ transformation. The relation~\eqref{star-triangle-C} underlies the integrability of the noncompact ${\rm SL}(2,\mathbb C)$ spin chain magnets.

In~\cite{Bazhanov:2007vg,Kels:2013ola,Kels:2015bda} new solutions of the star-triangle relation have been derived from the elliptic star-triangle relation~\cite{Bazhanov:2010kz,Derkachov:2012iv,Spiridonov18,Spiridonov:2010em}. Below we show that these star-triangle relations can be derived from the integrals~\eqref{Gustafson12}.

First we consider the relation associated with the integral~\eqref{gustafson1}. To this end we define the propagator
\begin{gather}
\boldsymbol S_\alpha(u) = (-1)^{[\alpha/2+u]}\boldsymbol\Gamma\left(\frac{1-\alpha}2 + u\right)
\boldsymbol\Gamma\left(\frac{1-\alpha}2 - u\right)\nonumber\\
\hphantom{\boldsymbol S_\alpha(u)}{}
=(-1)^{[\alpha/2+u]}\frac{\Gamma\left(\frac{1-\alpha}2 + u\right)}{\Gamma\left(\frac{1+ \bar \alpha}2 - \bar u\right)}
\frac{\Gamma\left(\frac{1-\alpha}2 - u\right)}
{\Gamma\left(\frac{1+\bar \alpha}2 + \bar u\right)}.\label{propS}
\end{gather}
The variables $u$ and $\alpha$ have the form
\begin{gather*}
u=n/2+{\rm i}\nu, \qquad \bar u=-n/2+{\rm i}\nu,\qquad \alpha=m +\sigma,\qquad \bar \alpha=-m+ \sigma ,
\end{gather*}
where $[u]= n$ and $\frac{1}{2}[\alpha]= m$ are either both integer or half-integer numbers and $\langle u \rangle =\nu$, $\langle \alpha \rangle = \sigma$ are real complex numbers, respectively. Under these conditions the arguments of the $\boldsymbol \Gamma$-functions in \eqref{propS} have the form~\eqref{def:z}. Slightly abusing the terminology we call the~$u$~($\alpha$) even or odd depending on the character of~$n$~($m$) and refer to this property as parity. Also, in order to avoid possible misunderstanding due to our agreement to indicate only the ``holomorphic'' arguments of functions, $f(\alpha)\equiv f(\alpha,\bar\alpha)$, we accept that, whenever $\bar x$ is not explicitly defined, $x+\alpha \equiv (x+\alpha,x+\bar\alpha)$, e.g., $\boldsymbol \Gamma(1/2+z)\equiv \boldsymbol\Gamma(1/2+z,1/2+\bar z)$.

The propagator $\boldsymbol S_\alpha$ inherits many properties of $\boldsymbol s_\alpha$:
\begin{itemize}\itemsep=0pt
\item for even(odd) $\alpha$ the propagator is an even(odd) function of $u$
\begin{gather*}
\boldsymbol S_\alpha(-u)=(-1)^{[\alpha]}\boldsymbol S_\alpha(u),
\end{gather*}
\item for imaginary $\langle \alpha \rangle = \sigma$, $\boldsymbol S_\alpha(u)(\boldsymbol S_\alpha(u))^\dagger=1$, i.e., the propagator reduces to a phase factor, while for $[\alpha]=0$ and $\langle \alpha \rangle$ real, $\boldsymbol S_\alpha(u)$ is real and positive.
\end{itemize}

The chain relation for the propagator $\boldsymbol S$ follows from the integral~\eqref{gustafson1} for $N=1$. Namely, making the substitution
\begin{alignat*}{3}
& z_1=(1-\alpha_1)/2+ z,\qquad && w_1=(1-\alpha_1)/2- z, & \\
& z_2=(1-\alpha_2)/2+ w, \qquad && w_2=(1-\alpha_2)/2- w&
\end{alignat*}
in \eqref{gustafson1} one obtains
\begin{gather}\label{Gamma-chain}
\sum_{n=-\infty}^\infty \int_{-{\rm i}\infty}^{{\rm i}\infty}
\boldsymbol S_{\alpha_1}(z-u)\boldsymbol S_{\alpha_2}(u-w) \frac{{\rm d}\nu}{2\pi {\rm i} } =\frac{\boldsymbol\Gamma(1-\alpha_1)\boldsymbol\Gamma(1-\alpha_2)}
{\boldsymbol\Gamma(2-\alpha_1-\alpha_2)} \boldsymbol S_{\alpha_1+\alpha_2-1}(z-w),
\end{gather}
where $u=n/2+\nu$ and sum goes over integers. The parity of the $\alpha_1$ and $z$ ($\alpha_2$ and $w$) is always the
same. The integral~\eqref{Gamma-chain} is well defined provided $\operatorname{Re}\langle \alpha_k\rangle <1$, $k =1,2$ and
$\operatorname{Re}\langle \alpha_1+\alpha_2\rangle > 1$: the poles of the $\boldsymbol\Gamma$ functions in the
integral~\eqref{Gamma-chain} are separated by the integration contour if $\operatorname{Re}\langle\alpha_k\rangle <1$ and the
integral converges at large $u$ if $\operatorname{Re}\langle \alpha_1+\alpha_2\rangle > 1$.

The star-triangle relation for $\boldsymbol S_\alpha$ can be obtained from the integral identity~\eqref{gustafson1red}
for $N=2$. Let us make the following substitution
\begin{gather*}
z_i\mapsto\frac{1-\alpha_i}2+z_i, \qquad \bar z_i\mapsto\frac{1-\bar \alpha_i}2+\bar z_i, \qquad w_i\mapsto\frac{1-\alpha_i}2-z_i,
\qquad \bar w_i\mapsto\frac{1-\bar \alpha_i}2-\bar z_i
\end{gather*}
in that equation. Taking into account that the condition $\sum_k (z_k+w_k)=1$ gives rise to the following restriction on the indices: $\sum_k\alpha_k=\sum_k\bar \alpha_k=2$, one derives after some algebra
\begin{gather}
 \sum_{n=-\infty}^\infty \int_{-{\rm i}\infty}^{{\rm i}\infty}
 \prod_{k=1}^3 \boldsymbol S_{\alpha_k}(z_k-u) \frac{{\rm d}\nu}{2\pi {\rm i} }\nonumber\\
 \qquad{} =\left[\prod_{k=1}^3 \boldsymbol \Gamma(1-\alpha_k)\right]
\boldsymbol S_{1-\alpha_1}(z_2-z_3)\boldsymbol
S_{1-\alpha_2}(z_3-z_1)\boldsymbol S_{1-\alpha_3}(z_1-z_2). \label{Gamma-star-triangle}
\end{gather}
Here, again, the $\alpha_k$ and $z_k$ are even or odd simultaneously. For the special choice of the parameters,
$\alpha_k=\bar\alpha_k$, this relation coincides with the star-triangle relation~\cite[equation~(22)]{Bazhanov:2007vg}.

Proceeding with the second integral~\eqref{gustafson2} we define the propagator as the product of four~$\boldsymbol\Gamma$ functions
\begin{gather*}
\boldsymbol {\mathcal D}_\alpha(z_1,z_2)= \boldsymbol\Gamma\left(\frac{1-\alpha}{2}\pm z_1 \pm z_2\right) =
\frac{\Gamma\left(\frac{1-\alpha}{2}\pm z_1 \pm z_2\right)}{\Gamma\left(
 \frac{1+\bar \alpha}{2}\pm \bar z_1 \pm \bar z_2\right)} .
\end{gather*}
The requirement for the arguments of $\boldsymbol\Gamma$ functions, $\frac{1-\alpha}{2}\pm z_1 \pm z_2$, to be integers imposes obvious restrictions on the relative parity of all variables. Obviously, the propagator $\boldsymbol {\mathcal D}_\alpha(z_1,z_2)$ is an even function of $z_1$, $z_2$ and invariant under the $z_1\leftrightarrow z_2$ permutation. Therefore, for each variable $z_k=n_k/2+\nu_k$ one can restrict $n_k$ to positive (negative) values. Note that unlike the previous case the pro\-pa\-ga\-tor $\boldsymbol {\mathcal D}_\alpha$ is not shift invariant. Also, up to a phase factor depending on the parity of arguments, $\boldsymbol {\mathcal D}_\alpha(z_1,z_2)\sim \boldsymbol S_\alpha(z_1-z_2)\boldsymbol S_\alpha(z_1+z_2)$.

The chain relation for the propagator $\boldsymbol {\mathcal D}_\alpha$ follows from the identity~\eqref{gustafson2}. Indeed, after the substitution $z_{1(2)}=(1-\alpha_1)/2\pm z$, $z_{3(4)}=(1-\alpha_2)/2\pm w$ the integral~\eqref{gustafson2} takes the form
\begin{gather*}
2\int_\pm \mathcal{D} u \|u\|^2 \boldsymbol {\mathcal D}_{\alpha_1}(z,u) \boldsymbol {\mathcal D}_{\alpha_2}(u,w) =
\pm \frac{\boldsymbol\Gamma(1-\alpha_1)\boldsymbol\Gamma(1-\alpha_2)}{
\boldsymbol\Gamma(2-\alpha_1-\alpha_2)}
\boldsymbol {\mathcal D}_{\alpha_1+\alpha_2-1}(z, w) .
\end{gather*}
Here the subscripts $\pm$ indicate that the sum goes over all integers (``$+$'') or half-integers (``$-$'') and we also recall that for $u=n/2+i \nu$, $\|u\|^2=\nu^2+n^2/4$.

Next, substituting $z_{2i-1}=(1-\alpha_i)/2+z_i$, $z_{2i}=(1-\alpha_i)/2-z_i$, for $i=1,2,3$ in equation~\eqref{gustafson2-red2} for $N=2$ one gets
\begin{gather}
2\int_\pm \mathcal{D} u \|u\|^2 \prod_{k=1}^3 \boldsymbol{\mathcal D}_{\alpha_k}(z_k,u) \nonumber\\
\qquad{} = \left(\prod_{k=1}^3 \boldsymbol \Gamma(1-\alpha_k)\right)
\boldsymbol{\mathcal D}_{1-\alpha_1}(z_2,z_3) \boldsymbol{\mathcal D}_{1-\alpha_2}(z_1,z_3)
\boldsymbol{\mathcal D}_{1-\alpha_3}(z_1,z_2), \label{stopen}
\end{gather}
where $\sum_k\alpha_k=\sum_k\bar\alpha_k=2$. As was mentioned earlier the parity of all variables have to be coordinated so that~\eqref{stopen} encompasses four different identities:
\begin{enumerate}\itemsep=0pt
\item[(1)] All $\alpha_k$ are integer / have positive parity
\begin{enumerate}\itemsep=0pt
\item[(a)] $z_k$, $u$ are even,
\item[(b)] $ z_k$, $u$ are odd,
\end{enumerate}
\item[(2)] $\alpha_1$ and $\alpha_2$, $\alpha_3$ are even and odd, respectively
\begin{enumerate}\itemsep=0pt
\item[(a)] $u$ and $z_1$ are even and $z_2$, $z_3$ are odd,
\item[(b)] $u$ and $z_1$ are odd and $z_2$, $z_3$ are even.
\end{enumerate}
\end{enumerate}
Variant~(1a) corresponds to the star-triangle relation obtained by Kels~\cite[equation~(17)]{Kels:2013ola}. An extension of the star-triangle relation of~\cite{Kels:2013ola} which follows from the relation~(\ref{gustafson2-red2}) with $N=2$ was also considered by Sarkissian and Spiridonov~\cite{Spir}.

In Cases (1a) and (1b) the functions $\boldsymbol {\mathcal D}_\alpha(z_1,z_2)$ are real if $0<\alpha=\bar\alpha<1$: $\boldsymbol {\mathcal D}_\alpha(z_1,z_2) >0$ for even $z_1$, $z_2$ (Case~(1a)) and $-\boldsymbol {\mathcal D}_\alpha(z_1,z_2) >0$ for odd $z_1$, $z_2$ (Case~(1b)). In both cases these functions can be interpreted as the Boltzmann weights of lattice integrable models, for more details see references~\cite{Kels:2013ola,Kels:2015bda}.

\section{Quasi-classical limit}\label{section:quasi}

Let us consider the identities~\eqref{Gamma-chain},~\eqref{Gamma-star-triangle} when the external variables become large. We replace the variables $z_k (w_k)$ in~\eqref{Gamma-chain},~\eqref{Gamma-star-triangle} by $L z_k=L(x_k+{\rm i}y_k)$ and take the limit $L\to\infty$. The variables $x_k$, $y_k$ can be considered in this limit as continuous ones, so that $z_k\in \mathbb C$, $\bar z_k=z_k^*$. It is easy to check that the leading contribution to the integrals~\eqref{Gamma-chain},~\eqref{Gamma-star-triangle} comes from the region where $u_k\sim L$, so that we replace $u_k\mapsto L u_k$ as well. Moreover in this limit $\boldsymbol S_\alpha$ turns into $\boldsymbol s_\alpha$:
\begin{gather}\label{Ss}
L^{2\langle\alpha\rangle}\boldsymbol S_\alpha(Lz) \underset{L\to\infty}{\mapsto} z^{-\alpha}\bar z^{-\bar\alpha} =\boldsymbol s_\alpha(z).
\end{gather}
The sum over $n$ in $\int \mathcal D u$ can be replaced by the integral so that
\begin{gather}\label{DtoD}
\frac1{L^2}\sum_{n=-\infty}^\infty \int_{-{\rm i}\infty}^{{\rm i}\infty} \frac{{\rm d} \nu}{2\pi {\rm i}} \underset{L\to\infty}{\mapsto} \frac1\pi \int {\rm d}^2 u,
\end{gather}
where $u=u_x+{\rm i}u_y$ and ${\rm d}^2z ={\rm d}u_x{\rm d}u_y$. Taking into account equations~\eqref{Ss} and~\eqref{DtoD} one checks that in the limit $L\to\infty$ equations~\eqref{Gamma-chain},~\eqref{Gamma-star-triangle} become the chain and the star-triangle relations, equations~\eqref{chain-C} and \eqref{star-triangle-C}, respectively.

In what follows we study the quasi-classical limit the integrals~\eqref{gustafson1} and \eqref{gustafson1red} for general~$N$. First of all we rewrite these identities in term of the propagator $\boldsymbol S_\alpha$. After the change of variables
\begin{gather*}
z_k\mapsto\frac{1-\alpha_k}2+z_k, \qquad w_k\mapsto\frac{1-\alpha_k}2-z_k, \qquad k=1,\ldots,N+1.
\end{gather*}
Equation~\eqref{gustafson1} takes the following form
\begin{gather*}
 \frac 1{N!} \int \mathcal{D} u_1\cdots \mathcal Du_{N}
 \prod_{1\leq i<j \leq N} \|u_i-u_j\|^2
 \prod_{k=1}^N\prod_{m=1}^{N+1} \boldsymbol S_{\alpha_m}(z_m-u_k)
 \\ \qquad{} =(-1)^{\sum\limits_{m=1}^N m [\alpha_{m+1}]}\frac{\prod\limits_{k=1}^{N+1}
 \boldsymbol \Gamma(1-\alpha_k)}{\boldsymbol\Gamma\left(N+1-\sum\limits_{k=1}^{N+1}\alpha_k\right)}
\prod_{1\leq i<k \leq N+1} \boldsymbol S_{\alpha_i+\alpha_k-1}(z_{i}-z_k) .
\end{gather*}
Similarly, the identity~\eqref{gustafson1red} can be represented as follows
\begin{gather*}
 \frac 1{(N-1)!} \int \mathcal{D} u_1\cdots \mathcal Du_{N-1} \prod_{1\leq i<j \leq N-1} \|u_i-u_j\|^2
\prod_{k=1}^N\prod_{m=1}^{N+1} \boldsymbol S_{\alpha_m}(z_m-u_k)
 \\ \qquad{} =(-1)^{\sum\limits_{m=1}^N m [\alpha_{m+1}]}
 \left( \prod_{k=1}^{N+1}\boldsymbol \Gamma(1-\alpha_k) \right)
\prod_{1\leq i<k \leq N+1} \boldsymbol S_{\alpha_i+\alpha_k-1}(z_{i}-z_k) ,
\end{gather*}
and where $\sum_k \alpha_k=\sum_k \bar \alpha_k=N$.

In the quasi-classical limit these identities are reduced to the following two-dimensional integrals with power functions
\begin{gather}
 \frac 1{N!} \int\prod_{1\leq i<j \leq N} |u_i-u_j|^2\prod_{k=1}^N\prod_{m=1}^{N+1}
 [z_m-u_k]^{-\alpha_m} {\rm d}^2u_1\cdots {\rm d}^2u_N \nonumber\\
 \qquad {}=\pi^N (-1)^{\sum\limits_{m=1}^N m [\alpha_{m+1}]}\frac{\prod\limits_{k=1}^{N+1}
 \boldsymbol \Gamma(1-\alpha_k)}{\boldsymbol\Gamma
 \left(N+1-\sum\limits_{k=1}^{N+1}\alpha_k\right)} \prod_{1\leq i<k \leq N+1}
[z_{i}-z_k]^{1-\alpha_i-\alpha_k}\label{LF}
\end{gather}
and
\begin{gather}
 \frac 1{(N-1)!} \int\prod_{1\leq i<j \leq N-1} |u_i-u_j|^2
 \prod_{k=1}^N\prod_{m=1}^{N+1}
 [z_m-u_k]^{-\alpha_m} {\rm d}^2u_1\cdots {\rm d}^2u_{N-1}
\nonumber \\ \qquad{} =\pi^{N-1}(-1)^{\sum\limits_{m=1}^N m [\alpha_{m+1}]}
 \left( \prod_{k=1}^{N+1}\boldsymbol \Gamma(1-\alpha_k) \right)
\prod_{1\leq i<k \leq N+1}
[z_{i}-z_k]^{1-\alpha_1-\alpha_k}.\label{LF1}
\end{gather}
It can be checked that in the quasi-classical limit the propagator $\boldsymbol{\mathcal D}_\alpha(z,w)$ turns into $
\big[z^2-w^2\big]^{-\alpha}. $ Therefore equations~\eqref{gustafson2} and equation~\eqref{gustafson2red} do not produce new identities
in this limit but reduce to the integrals~\eqref{LF} and \eqref{LF1} after an appropriate change of variables.

One can take a different point of view on the integrals \eqref{gustafson1} and \eqref{gustafson2} and consider them as the ``quantized'' version of the integrals~\eqref{LF}. For $\alpha_k=\bar\alpha_k$ the integral~\eqref{LF} is the special case of the duality relation~\cite{Baseilhac:1998eq} for the Dotsenko--Fateev (DF) integrals~\cite{Dotsenko:1984ad} considered in the next section. Therefore one can hope that there exists a~``quantized'' version of the duality relation in a general situation.

\section{Dotsenko--Fateev integrals}\label{sect:FLI}

In this section we give an elementary proof of the following relation
\begin{gather}
\frac{1}{\pi^n n!}\int
\prod_{i<k}^{n} [y_i-y_k]
\prod_{i=1}^n \prod_{j=1}^{n+m+1} [y_i-z_j]^{-\alpha_j}
{\rm d}^2y_1\cdots {\rm d}^2y_n
 =
\frac{ (-1)^{\sum\limits_{k=1}^{n+m} k[\alpha_{k+1}]}
\prod\limits_{j=1}^{n+m+1} \boldsymbol\Gamma(1-\alpha_j)}{\boldsymbol\Gamma\left(1+n -\sum\limits_{j=1}^{n+m+1} \alpha_j\right)}\nonumber\\
\times\prod_{i<j}^{n+m+1} [z_j-z_i]^{1-\alpha_i-\alpha_j} \frac1{\pi^m m!}
\int\prod_{i<k}^{m} [u_i-u_k] \prod_{i=1}^m \prod_{j=1}^{n+m+1} [u_i-z_j]^{-1+\alpha_j} {\rm d}^2u_1\cdots {\rm d}^2u_m ,\label{duality}
\end{gather}
which is, provided the parameters satisfy the constraint $\alpha_k=\bar\alpha_k$, the duality
relation~\cite{Baseilhac:1998eq,Fateev:2007qn} for the DF integrals~\cite{Dotsenko:1984ad}, see
also~\cite{Pasquetti:2019tix}. For $m=0$ this identity coincides with~\eqref{LF}.

To evaluate the integral on the l.h.s.\ of~\eqref{duality} we go over to the variables $x_k=x_k(y_1,\ldots,y_n)$ which are essentially the elementary symmetric functions
\begin{gather*}
\prod_{i=1}^n (y_i+t) = x_1+x_2 t+x_3 t^2 +\cdots +x_{n} t^{n-1} + t^n.
\end{gather*}
Every point in $x$-space has $n!$ preimages in $y$-space and the Jacobian of the transformation is $J = \big|\frac{\partial x_k}{\partial y_i}\big|^2 = \prod\limits_{i<j}^{n} [y_i-y_k]$. Introducing the notation $t_j=-z_j$ one gets for the integral on the l.h.s.\ of~\eqref{duality}
\begin{gather}\label{eq:lhs}
\pi^{-n} \int\prod_{j=1}^{n+m+1}
\big[x_1+x_2 t_j+x_3 t_j^2 +\cdots +x_{n}t_j^{n-1} +t_j^n\big]^{-\alpha_j}
{\rm d}^2 x_1\cdots {\rm d}^2 x_{n}.
\end{gather}
Then using the momentum space representation for the propagators
\begin{gather*}
[z]^{-\alpha}=\frac1{\pi} {\rm i}^{\alpha-\bar\alpha} \boldsymbol\Gamma(1-\alpha)
\int {\rm e}^{{\rm i}(pz+\bar z\bar p)} [p]^{\alpha-1} {\rm d}^2p
\end{gather*}
in~\eqref{eq:lhs} and carrying out all integrals over $x_k$ one gets
\begin{gather*}
\frac{ {\rm i}^{\sum_j (\alpha_j-\bar \alpha_j)}}{\pi^{m+1}} \prod_{k=1}^{n+m+1} \boldsymbol\Gamma(1-\alpha_k)
\prod_{i=1}^{n+m+1} \int {\rm d}^2p_i [p_i]^{\alpha_k-1} \prod_{k=1}^{n}\delta^{(2)}\left(\sum_j p_j t_j^{k-1}\right)
{\rm e}^{{\rm i}\sum_j ( p_j t_j^n+ \bar p_j \bar t_j^n)}\\
= \frac{\prod\limits_{k=1}^{n+m+1} \boldsymbol\Gamma(1-\alpha_k)}{\pi^m\boldsymbol\Gamma\left(n+1-\sum_j \alpha_j\right)}
 \prod_{i=1}^{n+m+1} \int
\frac{{\rm d}^2p_i}{ [p_i]^{1-\alpha_i}} \prod\limits_{k=1}^{n}\delta^{(2)}\left(\sum_j p_j t_j^{k-1}\right)
\delta^{(2)}\left(\sum_j p_j t_j^{n}-1\right).
\end{gather*}
In order to get the last $\delta$-function one represents ${\rm e}^{{\rm i}\sum_j ( p_j t_j^n+ \bar p_j \bar t_j^n)}$ as
$\int {\rm e}^{{\rm i}(\lambda+\bar\lambda)}\delta^{(2)}(\lambda-\sum_j p_j t_j^n) d\lambda$ and rescales $p_j\to\lambda p_j$. The delta
functions cut out a $m$-dimensional subspace in the $(n+m+1)$-dimensional space defined by the linear equations
\begin{gather}\label{sys}
\sum_{j=1}^{n+m+1} p_j t_j^{k-1}=0, \qquad k=1,\ldots, n \qquad \text{and} \qquad \sum_{j=1}^{n+m+1} p_j t_j^{n}=1 .
\end{gather}
The solution depends on $m$ free variables, $u_1,\ldots, u_m$, and takes the form
\begin{gather}\label{pj}
p_j(u_1,\ldots,u_m) = \lambda_j (u_1+t_j)\cdots(u_m+t_j),
\end{gather}
where $\lambda_j=\prod\limits_{k\neq j} (t_j-t_k)^{-1}$, $j=1,\ldots,n+m+1$. That~$p_j$, \eqref{pj}, satisfy equations~\eqref{sys} follows from an identity due to Milne \cite[Lemma~1.33]{MR812934}, see also~\cite[equation~(2.3)]{MR2026866},
\begin{gather}\label{Milne}
\sum_{k=1}^N \frac{(b_1-t_k)\cdots(b_N-t_k)} {t_k\prod\limits_{j\neq k}(t_j-t_k)} = \frac{b_1\cdots b_N}{t_1\cdots t_N} -1 .
\end{gather}
To see it is enough to put $N=m+n+1$, $b_k =-u_k$ for $k\leq m$ and $b_{k}=-u$, for $k>m$ in the above identity and compare the coefficients at the powers $u^k,$ $k=0,\ldots,n$ on both sides of \eqref{Milne}.

Making the change of variables
\begin{alignat*}{3}
&p_j= \lambda_j (u_1+t_j)\cdots(u_m+t_j) +s_j,\qquad && j=1,\ldots,n+1, &\\
&p_j=\lambda_j (u_1+t_j)\cdots(u_m+t_j),\qquad && j=n+2,\ldots, n+m+1,&
\end{alignat*}
and taking into account that
\begin{gather*}
 \prod_{k=1}^{n}\delta^2\left(\sum_{j=1}^{n+m+1} p_j t_j^{k-1}\right)\delta^2
\left(\sum_{j=1}^{n+m+1} p_j t_j^{n}-1\right) =
 \prod_{k=1}^{n+1}\delta^2\left(\sum_{j=1}^{n+1} s_j t_j^{k-1}\right) =
\frac{\prod\limits_{j=1}^{n+1}\delta^2\left(s_j\right)}
{\prod\limits_{1\leq i<j\leq n+1} [t_i-t_j]}
\end{gather*}
and
\begin{gather*}
\prod_{j=1}^{n+m+1} {\rm d}^2 p_j =\left(
 \prod_{n+2\leq i<j\leq n+m+1}[t_i-t_j] \prod_{k=n+2}^{n+m+1}[\lambda_{k}]\right)
\frac1{m!}\prod_{j=1}^{m} {\rm d}^2 u_j \prod_{j=1}^{n+1} {\rm d}^2 s_j
\end{gather*}
one gets after some algebra
\begin{gather*}
\frac{\prod\limits_{k=1}^{n+m+1} \boldsymbol\Gamma(1-\alpha_k)}{\boldsymbol\Gamma\left(n+1-\sum_j \alpha_j\right)}
\left(\prod_{j<k}(-1)^{\alpha_k-\bar\alpha_k}[t_j-t_k]^{1-\alpha_k-\alpha_j}\right)\\
\qquad{} \times \frac1{\pi^m m!}\prod_{j=1}^{n+m+1}\int\prod_{k=1}^m (u_k+t_j)^{\alpha_j-1}{\rm d}^2u_1\cdots {\rm d}^2u_{m}.
\end{gather*}
It coincides with the r.h.s.\ of equation~\eqref{duality} after changing $t_j\to -z_j$. We have learned from the discussions with Litvinov that similar proof of the duality relation~\eqref{duality} is presented in his lectures on conformal field theory~\cite{Litvinov}.

Finally, we make a conjecture that the duality relation~\eqref{duality} admits a generalization to the ``quantized'' cases. Namely, these relations take the form
\begin{gather}
 \frac1{n!} \int\mathcal{D}u_1 \cdots\mathcal{D}u_n \prod_{i<j}\|u_i-u_j\|^2
 \prod_{i=1}^n \prod_{j=1}^{n+m+1} (-1)^{[u_i]}
\boldsymbol\Gamma(z_j-u_i)\boldsymbol\Gamma(u_i+w_j)\nonumber\\
\qquad{} =\frac{\prod\limits_{i,j=1}^{n+m+1}\boldsymbol \Gamma(z_i+w_j)}{
\boldsymbol\Gamma\left(\sum_j (z_j+w_j)-m\right)} (-1)^{m \sum_j [z_j-w_j]}
\frac1{m!} \int\mathcal{D}u_1 \cdots\mathcal{D}u_m \nonumber\\
\qquad\quad{}\times \prod_{i<j}\|u_i-u_j\|^2 \prod_{i=1}^m \prod_{j=1}^{n+m+1} (-1)^{[u_i]}
\boldsymbol\Gamma(z'_j-u_i)\boldsymbol\Gamma(w'_j+u_i),\label{Dual-1}
\end{gather}
where $z'_j=1/2-w_j$, $w'_j=1/2-z_j$ and
\begin{gather}
 \frac{2^n}{n!} \int_\pm
\mathcal{D}u_1 \cdots\mathcal{D}u_n \|u_k\|^2 \prod_{1\leq i<j\leq n}\|u_i\pm u_j\|^2
\prod_{i=1}^n \prod_{j=1}^{2(n+m+1)}
\boldsymbol\Gamma(z_j\pm u_i)\nonumber\\
\qquad{} =\varkappa_{n+m}\frac{\prod\limits_{i<j}^{2(n+m+1)}\boldsymbol \Gamma(z_i+z_j)}{
\boldsymbol\Gamma\left(\sum_j z_j-m\right)} \frac{2^m}{m!} \int_\pm\mathcal{D}u_1 \cdots\mathcal{D}u_m \|u_k\|^2 \nonumber\\
\qquad\quad{}\times \prod_{1\leq i<j\leq m}\|u_i\pm u_j\|^2 \prod_{i=1}^m \prod_{j=1}^{2(n+m+1)}\boldsymbol\Gamma(1/2-z_j\pm u_i),\label{Dual-2}
\end{gather}
where $\varkappa_k=1,(-1)^{k(k+1)/2}$ for the integer and half-integer cases, respectively. For $m=0$ these integrals
are equivalent to the integrals~\eqref{Gustafson12} and in the quasi-classical limit both of them reproduce the
duality relation~\eqref{duality}. For $n=m=1$ the relations~\eqref{Dual-1} and \eqref{Dual-2} follow from the
star-triangle relations~\eqref{Gamma-star-triangle} and \eqref{stopen}. For few first $n$ and $m$ the
integrals~\eqref{Dual-1},~\eqref{Dual-2} go through numerical tests. Closing this section we note that quite similar
duality relations were observed recently in the so-called conformal fishnet model~\cite{Derkachov:2018rot}.

\section{Summary}\label{sect:summary}

In~\cite{Derkachov:2016ucn,Derkachov:2017pvx} a generalization of Gustafson integrals to the complex case
have been obtained. The derivation of these integrals rely on the completeness of the SoV representation for the
${\rm SL}(2,\mathbb C)$ magnets, that is not yet proven. In this work we presented a direct calculation of two
$\boldsymbol\Gamma$ function integrals. We expect that these integral identities will be helpful in proving the
completeness of the SoV representation for the ${\rm SL}(2,\mathbb C)$ spin chains.

The complex integrals are, up to appropriate modification of the $\Gamma$ function and integration measure, exact
copies of the integrals obtained by R.A.~Gustafson~\cite{Gustafson94}. However, the analytic properties of these
integrals are different. We have shown that several, apparently distinct in the $\mathrm{SL}(2,\mathbb R)$ context
integrals are intrinsically related to each other in the $\mathrm{SL}(2,\mathbb C)$ formulation.

\looseness=-1 We have also shown that the complex $\boldsymbol\Gamma$ integrals for the lowest $N$ underlie the star-triangle
identities derived in references~\cite{Bazhanov:2007vg,Kels:2013ola,Kels:2015bda} and in the quasi-classical limit are
reduced to the special $(m=0)$ case of the duality relation for the DF
integrals~\cite{Baseilhac:1998eq,Dotsenko:1984ad,Fateev:2007qn}. We also conjecture an appropriate modification of
these duality relations to hold for the integrals with complex $\boldsymbol \Gamma$ functions.

\subsection*{Acknowledgements}

We are grateful to A.V.~Litvinov and V.P.~Spiridonov for fruitful discussions. {We would also like to express our sincere gratitude to the referees for their numerous useful comments and suggestions.} This study was supported by the Russian Science Foundation project No 19-11-00131 and Deutsche Forschungsgemeinschaft (A.M.), grant MO 1801/1-3.

\pdfbookmark[1]{References}{ref}
\LastPageEnding

\end{document}